\documentclass[11pt,a4paper]{article} 
\pdfoutput=1
\usepackage{jcappub}
\newcommand{\eq}[1]{\begin{equation}#1\end{equation}}
\newcommand{\eqa}[1]{\begin{eqnarray}#1\end{eqnarray}}
\newcommand{\secs}[1]{\section{#1\label{sec-#1}}}
\newcommand{\ssecs}[1]{\subsection{#1\label{ssec-#1}}}

\newcommand{\fig}[4]{\begin{figure}[#4]\centering\includegraphics[width=#3\textwidth]{Graph-#1.pdf}\caption{#2}\label{fig-#1}\end{figure}}

\newcommand{\refeq}[1]{eq.\ (\ref{eq-#1})}
\newcommand{\refsec}[1]{section \ref{sec-#1}}

\newcommand{\refig}[1]{figure \ref{fig-#1}}

\newcommand{\subs}[1]{_\mathrm{#1}}

\newcommand{\dd}[1]{\mathrm{d}#1}

\newcommand{\cm}[1]{}
\newcommand{\vect}[1]{\mathbf{#1}}
\newcommand{\vectf}[1]{_{\vect{#1}}}
\newcommand{\direc}[1]{\hat{\vect{#1}}}

\def\fNL{f\subs{NL}}
\def\tNL{\tau\subs{NL}}

\def\ml{|_{sr}}

\begin{document}
\title{CMB dipole asymmetry from a fast roll phase}
\author{Anupam Mazumdar}
\author{and Lingfei Wang}
\affiliation{Physics Department, Lancaster University, Lancaster LA1 4YB, UK}
\abstract{The observed CMB (cosmic microwave background) dipole asymmetry cannot be explained by a single field model of inflation - it inevitably requires more than one field where one of the fields is responsible for amplifying the super-Hubble fluctuations beyond the pivot scale. Furthermore the current constraints on $\fNL$ and $\tNL$ require that such an amplification cannot produce large non-Gaussianity. In this paper we propose a model to explain this dipole asymmetry from a spectator field, which is responsible for generating all the curvature perturbations, but has a temporary fast roll phase before the Hubble exit of the pivot scale. The current data prefers spectator scenario because it leaves no isocurvature perturbations. The spectator model will also satisfy the well-known constraints arising from quasars, and the quadrupole and octupole of the CMB.}
\maketitle

\secs{Introduction}

The primordial inflation matches all the known predictions for the  cosmic microwave background (CMB) temperature anisotropy~\cite{Bennett:1996ce,Hinshaw:2012fq,Ade:2013zuv}. However recent observations from WMAP~\cite{Eriksen:2003db,Eriksen:2007pc} and Planck~\cite{Ade:2013nlj} found a dipole asymmetry in the CMB power spectrum with $3\sigma$ significance. The origin of this asymmetry is hard to explain within a single field model of inflation. It was first pointed out in Refs.~\cite{Erickcek:2008sm,Erickcek:2009at} that an initial perturbation with a scale larger than the CMB pivot scale could be responsible for the asymmetry by favoring a certain direction. This would inevitably require more than one field dynamics during inflation, because a single field inflation can not give rise to a large non-Gaussianity required to explain the dipole asymmetry \footnote{One has to further make sure that there is no residual isocurvature perturbations~\cite{Ade:2013uln}, and the end of inflation {\it solely} excites the Standard Model relativistic degrees of freedom \cite{Wang:2013hva}, since there is no evidence of dark radiation from Planck \cite{Ade:2013zuv}. This already constrains models of inflation, which can explain both CMB perturbations and ensure right thermal history of the universe \cite{Mazumdar:2010sa}. Only visible sector models of inflation, where inflation is embedded within a Standard Model gauge theory or its minimal supersymmetric extensions can ensure a consistent scenario for a successful inflationary paradigm, see Refs.~\cite{Allahverdi:2006iq,Allahverdi:2006we,Allahverdi:2006cx,Choudhury:2013jya} and \cite{Wang:2013hva}.}.

It is also possible that there could be more fields during inflation, which are responsible for 
creating {\it large }non-Gaussianity with opposite signs, i.e. $\pm \fNL$~\cite{Wang:2013lda}. Their fine cancellations could yield not only the dipole asymmetry, but would also match the current limits on $\fNL=2.7\pm11.6$ (at $2\sigma$). Similarly there are other interesting suggestions, see~\cite{Dai:2013kfa,Lyth:2013vha,Chang:2013vla,McDonald:2013aca,Namjoo:2013fka,Zhao:2013jya}. Other attempts have also been made to address multiple CMB anomalies altogether in \cite{Donoghue:2007ze,Chen:2013eaa,Liu:2013kea}.

As it was pointed out in Ref.~\cite{Wang:2013lda}, one would require a violation of slow-roll dynamics for the fields responsible for creating super-long wavelength perturbations. One simple way of obtaining this would be via a brief period of {\it fast roll} phase~\cite{Linde:2001ae}, before the relevant perturbations have left the Hubble patch. The fast roll phase enhances the perturbations on scales larger than the pivot scale, but leaves the pivot and smaller scales unaffected. However, one has to check whether the dipole asymmetry can contaminate the quadrupole and the octupole. 

In this paper we will study all these constraints within a simple scenario where we have inflaton within a visible sector, and a spectator field which decays during inflation but after sourcing the large scale fluctuations, as discussed in Refs.~\cite{Mazumdar:2012rs,Wang:2013oea}
\footnote{In principle a curvaton scenario \cite{Moroi:2001ct,Lyth:2001nq,Enqvist:2001zp,Lyth:2002my} could also provide the dipole asymmetry within our mechanism, but since the curvaton decays after inflation, it can potentially create large non-Gaussianity and large isocurvature fluctuations. Furthermore, one has to ensure that the curvaton decays into the standard model degrees of freedom \cite{Mazumdar:2011xe,Wang:2013hva}. In this respect a spectator field which decays during inflation could be much more helpful for model building \cite{Wang:2013hva,Mazumdar:2012rs,Wang:2013oea}, because it does not leave any isocurvature fluctuations, and naturally provides smaller non-Gaussianity than the curvaton, while the inflaton could be within a visible sector. }. 

We give a brief overview on the CMB dipole asymmetry in \refsec{CMB dipole asymmetry}. We study the dipole asymmetry for a slow roll spectator field in \refsec{Slow roll scenario}. The enhancement in the perturbations leading to the dipole asymmetry has been computed in \refsec{Tachyonic fast roll scenario}. We consider the known constraints in \refsec{Known constraints}, and provide a viable example in \refsec{A viable model}. We conclude in \refsec{Conclusion}.

\secs{CMB dipole asymmetry}
The dipole asymmetry in the CMB power spectrum can be modeled by the directional dependence $\direc{n}$ in the temperature fluctuations, given by\footnote{
We use bold symbols to indicate vectors.}
(see for instance \cite{Eriksen:2007pc})
\eq{\delta T(\direc{n})=(1+A\,\direc{p}\cdot\direc{n})\overline{\delta T}(\direc{n}),}
where the dipole asymmetry is along the direction $\direc{p}$, and its strength is $0<A\ll1$. After extracting the dipole asymmetry, $\overline{\delta T}(\direc{n})$ then has an isotropic Gaussian distribution. If we pick a local patch in the direction $\direc{n}$ on the CMB map, and calculate the CMB power spectrum only in this patch, it will also acquire a directional dependence (neglecting ${\cal O}(A^2)$),
\eq{P_{\delta T}(\direc{n})=(1+2A\,\direc{p}\cdot\direc{n})P_{\overline{\delta T}}.}
This directional dependence becomes most significant when we compare the two opposite directions, $\direc{n}=\direc{p}$ and $\direc{n}=-\direc{p}$. Their relative difference is given by
\eq{\frac{P_{\delta T}(\direc{p})-P_{\delta T}(-\direc{p})}{P_{\overline{\delta T}}}=4A.\label{eq-ca-PTd}}

The latest Planck observations \cite{Ade:2013nlj} constrain the value, $A=0.07\pm0.02$, while confirming the previous analysis on the WMAP data \cite{Eriksen:2003db,Eriksen:2007pc}. Since the temperature anisotropy is seeded by the primordial curvature perturbations, it is straightforward to think that its dipole asymmetry may share the same origin.

The dipole asymmetry may arise from a scalar field, $\sigma$, which partly or totally contributes to the primordial curvature perturbations, and has a non-uniform background value at the Hubble exit. This can be caused by the initial field perturbations of $\sigma$, whose wavelengths are much larger than that of the pivot scale, as argued in Refs.\ \cite{Erickcek:2008sm,Erickcek:2009at,Lyth:2013vha,Wang:2013lda,McDonald:2013aca,Namjoo:2013fka}.

Therefore, we can assume such initial field perturbations of $\sigma$ can create the difference $\Delta\sigma$ in its background evolutions of the two local universe patches  along the $\direc{p}$ and $-\direc{p}$ directions. This will result in the difference in the power spectrum of the primordial curvature perturbations, $\zeta$, by an amount
\eq{\Delta P_\zeta=\frac{\partial P_\zeta}{\partial\sigma}\Delta\sigma.}
Since $P_{\delta T}\propto P_\zeta$ at the linear order, from \refeq{ca-PTd} we know
\eq{A=\frac{\Delta P_\zeta}{4P_\zeta}=\frac{1}{4}\frac{\partial P_\zeta}{P_\zeta\partial\sigma}\Delta\sigma.\label{eq-ca-Ads}}

We can write $A$ as a function of the power spectrum of the curvature perturbations $P_\zeta$ and that of the field perturbations $P_{\delta\sigma_*}$, as well as the primordial local bispectrum $\fNL$ or trispectrum $\tNL$, see \cite{Lyth:2013vha,Wang:2013lda}. There are two ways of satisfying small $\fNL$ and the observed dipole asymmetry.

\begin{itemize}
\item {\bf Inflation and the spectator}: If $\sigma$ is a spectator field, which is the only source of the primordial curvature perturbations, the amount of CMB dipole asymmetry can be shown as~\cite{Lyth:2013vha,Wang:2013lda}:
\eq{A=\frac{3}{5}\frac{|\Delta\sigma|}{\sqrt{P_{\delta\sigma_*}}}\left|\fNL\right|\sqrt{P_\zeta}.\label{eq-ca-Aeqs}}
The latest Planck observations \cite{Ade:2013ydc,Ade:2013nlj,Ade:2013zuv} give the central values $P_\zeta=2.196\times10^{-9}$, $A=0.07$. The local bispectrum is constrained by $\fNL=2.7\pm5.8$, which can provide an upper bound as $|\fNL|<14.3$ ($@>95\%$ C.L.). To achieve $A=0.07$, we would require:
\eq{\frac{|\Delta\sigma|}{\sqrt{P_{\delta\sigma_*}}}>174.\label{eq-ca-Dslims}}

\item {\bf Inflaton, spectator, and the other}: When the spectator $\sigma$ coexists with other sources of curvature perturbations, then they may generate opposite local bispectra $\pm \fNL$ ( individual contributions could be large ), which mostly cancel to yield a small total $\fNL$. The negative values of $\fNL$ may come from
preheating~\cite{Enqvist:2004ey,Enqvist:2005qu,Jokinen:2005by}. 
This can enhance the CMB dipole asymmetry, to the maximum extent as \cite{Wang:2013lda}
\eq{A<\frac{|\Delta\sigma|}{2\sqrt{P_{\delta\sigma_*}}}\sqrt{\tNL P_\zeta}.\label{eq-ca-Aup}}
From the $95\%$ CL upper bound on $\tNL<2800$, we only need a weaker field difference, $\Delta\sigma$, in order to generate the CMB dipole asymmetry $A=0.07$, as
\eq{\frac{|\Delta\sigma|}{\sqrt{P_{\delta\sigma_*}}}>56.\label{eq-ca-Dslim}}
\end{itemize}

In this paper we will be focusing on how to excite the $\sigma$ field such that one can obtain large values of overall $\Delta \sigma$.

\section{Slow roll evolution with two fields\label{sec-Slow roll scenario}}
It has been briefly discussed in \cite{Wang:2013lda} that if the slow roll approximations are well satisfied for a curvaton or spectator field $\sigma$, its initial perturbations of any single wavelength would not give rise to any significant CMB asymmetry.

In the simplest scenario, we can think of a specator field $\sigma$ as the source of curvature perturbations. The inflation is dominated by another field $\phi$, which rolls very slowly, providing a nearly constant Hubble rate of expansion $H$ during inflation. For minimalism, we assume $\sigma$ has negligible interactions with the inflaton, so any perturbation in $\sigma$ will not affect the inflaton, and will only convert to curvature perturbations well after the corresponding Hubble exit. Its energy density should also be subdominant than that of the inflaton's. The action is given by:
\eq{S=\int\dd^4x\sqrt{|-g_{\mu\nu}|}\left(-\frac{1}{2}\partial_\mu\sigma\partial^\mu\sigma-\frac{1}{2}m^2\sigma^2+{\cal L}_\phi\right),\label{eq-ml-action0}}
where $g_{\mu\nu}=\mathrm{diag}(-1,a^2,a^2,a^2)$ with $a(t)$ being the scale factor, and ${\cal L}_\phi$ is the Lagrangian density for the inflaton, which may arise from the visible sector. We assume $m \ll H$ during inflation.

\ssecs{Fluctuations of a light field}
We can separate the inhomogeneous part of $\sigma$ as its perturbations $\delta\sigma(x,t)$, by writing $\sigma(x,t)=\sigma(t)+\delta\sigma(x,t)$. From the total action \refeq{ml-action0}, the perturbation $\delta\sigma(x,t)$ yields to its equation of motion (assuming $m^2\ll H^2$)
\eq{\ddot{\delta\sigma}+3H\dot{\delta\sigma}-\partial_i\partial^i\delta\sigma=0,}
where dot means taking derivative w.r.t time $t$. A Fourier transformation into the momentum space $\delta\sigma\vectf k(t)$ then gives
\eq{\ddot{\delta\sigma}\vectf k+3H\dot{\delta\sigma}\vectf k+\frac{k^2}{a^2}\delta\sigma\vectf k=0.\label{eq-ml-dsk}}

For the sub-Hubble modes with $k^2\gg a^2H^2$, we define the conformal time $\dd\tau\equiv\dd t/a$, and $u\vectf k\equiv a\delta\sigma\vectf k$. We use prime to indicate $\dd/\dd\tau$, and the equation of motion \refeq{ml-dsk} can be rewritten as
\eq{u\vectf k''+(k^2-2a^2H^2)u\vectf k=0,\hspace{0.3in}\mathrm{for\ }k^2\gg a^2H^2.\label{eq-ml-ueom}}
For the super-Hubble modes with $k^2\ll a^2H^2$, we similarly define $\psi\vectf k\equiv a^\frac{3}{2}\sigma\vectf k$. This reduces \refeq{ml-dsk} to
\eq{\ddot\psi\vectf k-\left(\frac{9}{4}H^2-\frac{k^2}{a^2}\right)\psi\vectf k=0,\hspace{0.3in}\mathrm{for\ }k^2\ll a^2H^2.\label{eq-ml-psieom}}
We do not specify the inflation model, but instead just assume $H$ remains constant throughout inflation. Then the universe will expand exponentially, but this will only make slow changes in the corresponding effective masses of \refeq{ml-ueom} and \refeq{ml-psieom}. The above solutions can be recast into
\eq{\langle|\delta\sigma\vectf k(N)|^2\rangle=e^{-2\int_{N_0}^N\alpha\vectf k(N)\dd N}\langle|\delta\sigma\vectf k(N_0)|^2\rangle,\label{eq-ml-ds}}
where $N\equiv\ln a$ is the number of e-folds of universe expansion, which we use as the proper time. After defining $N_k$ as
\eq{k^2=2e^{2N_k}H^2,}
we can write  $\alpha\vectf k(N)$ as
\eq{\alpha\vectf k(N)\ml=\left\{\begin{array}{l@{\hspace{0.5in}\mathrm{for\ }}l}
1,&N_k\ge N,\ (k^2\ge2a^2H^2)\\
\frac{3}{2}-\sqrt{\frac{9}{4}-2e^{2(N_k-N)}},&N_k<N,\ (k^2<2a^2H^2),
\end{array}\right.\label{eq-ml-as2}}
where we use the subscript ``$sr$'' to indicate the slow roll case.

\subsection{Lack of dipole asymmetry\label{ssec-CMB asymmetry deficiency}}
From \refeq{ml-as2}, we see that  as the universe expands, any mode $\vect k$ will reach $N\gg N_k$, where $\alpha\vectf k\ml\rightarrow0$. This means the perturbations $\delta\sigma\vectf k$ will gradually freeze. Based on the solution \refeq{ml-as2}, the canonical quantization then gives the simple relation $\langle|\delta\sigma\vectf k(N)|^2\rangle\ml\propto k^{-3}$ for $N\gg N_k$. This corresponds to the perturbation in the $x$ space
\eq{\langle|\delta\sigma(x)|^2\rangle=\int\langle|\delta\sigma\vectf k|^2\rangle\dd^3\vect k=\int P_{\delta\sigma_k}\dd\ln k,}
and here we have a scale invariant power spectrum $P_{\delta\sigma\vectf k}\ml\propto k^0$.

The perturbations will also generate a gradient along any arbitrary $z$ direction
\eq{\left\langle\biggl|\frac{\partial}{\partial z}\delta\sigma(x)\biggr|^2\right\rangle=\int\langle|\delta\sigma\vectf k|^2\rangle k_z^2\dd^3\vect k=\int P_{\delta\sigma_k}\frac{k^2}{2}\dd\ln k.}
This gradient will source the field difference between the opposite sides on the last scattering surface along the $z$ direction, by an amount
\eq{\Delta\sigma^2=4r\subs{ls}^2\left\langle\biggl|\frac{\partial}{\partial z}\delta\sigma(x)\biggr|^2\right\rangle,}
where $r\subs{ls}$, our distance to the last scattering surface, is defined by
\eq{r\subs{ls}=\frac{1}{a_*H_*}=\frac{e^{-N_*}}{H}.\label{eq-ml-rls}}
Therefore, the relative strength of the field difference becomes
\eq{\left.\frac{\Delta\sigma^2}{P_{\delta\sigma_*}}\right\ml=\int2r\subs{ls}^2k^2\dd\ln k=\int4e^{2(N_k-N_*)}\dd N_k.\label{eq-ml-Ds2}}

Note that only the scales much larger than the pivot scale can significantly contribute to the field difference $\Delta\sigma$ from their gradient contribution. This means the upper bound of the integral in \refeq{ml-Ds2} should be $N_k<N_*$. For this reason, the exponential suppression inside the integral will give rise to
\eq{\left.\frac{\Delta\sigma^2}{P_{\delta\sigma_*}}\right\ml=2e^{2(N_k-N_*)}<2.}
This is insufficient to generate the observed CMB asymmetry, see \refeq{ca-Dslim}. We conclude that a light canonical slow roll $\sigma$ field cannot generate sufficient CMB asymmetry.

\section{Tachyonic fast roll phase of a spectator\label{sec-Tachyonic fast roll scenario}}
As we suggested earlier in Ref.\ \cite{Wang:2013lda}, a violation in the slow roll conditions may produce sufficient CMB dipole asymmetry. It is well known that the perturbations can get enhanced during a tachyonic fast roll phase \cite{Linde:2001ae}. Therefore we wish to investigate how a tachyonic fast roll phase may enhance the CMB asymmetry.

Let us assume that the spectator field acquires a tachyonic mass for a brief period. The total action for the tachyonic phase can be given by
\eq{S=\int\dd^4x\sqrt{|-g_{\mu\nu}|}\left(-\frac{1}{2}\partial_\mu\sigma\partial^\mu\sigma+\frac{1}{2}m^2\sigma^2+{\cal L}_\phi\right),\label{eq-tc-action}}
where we can also define $N_m$ to write $m$ w.r.t $H$ as
\eq{m^2=(e^{2N_m}-1)2H^2.}

\fig{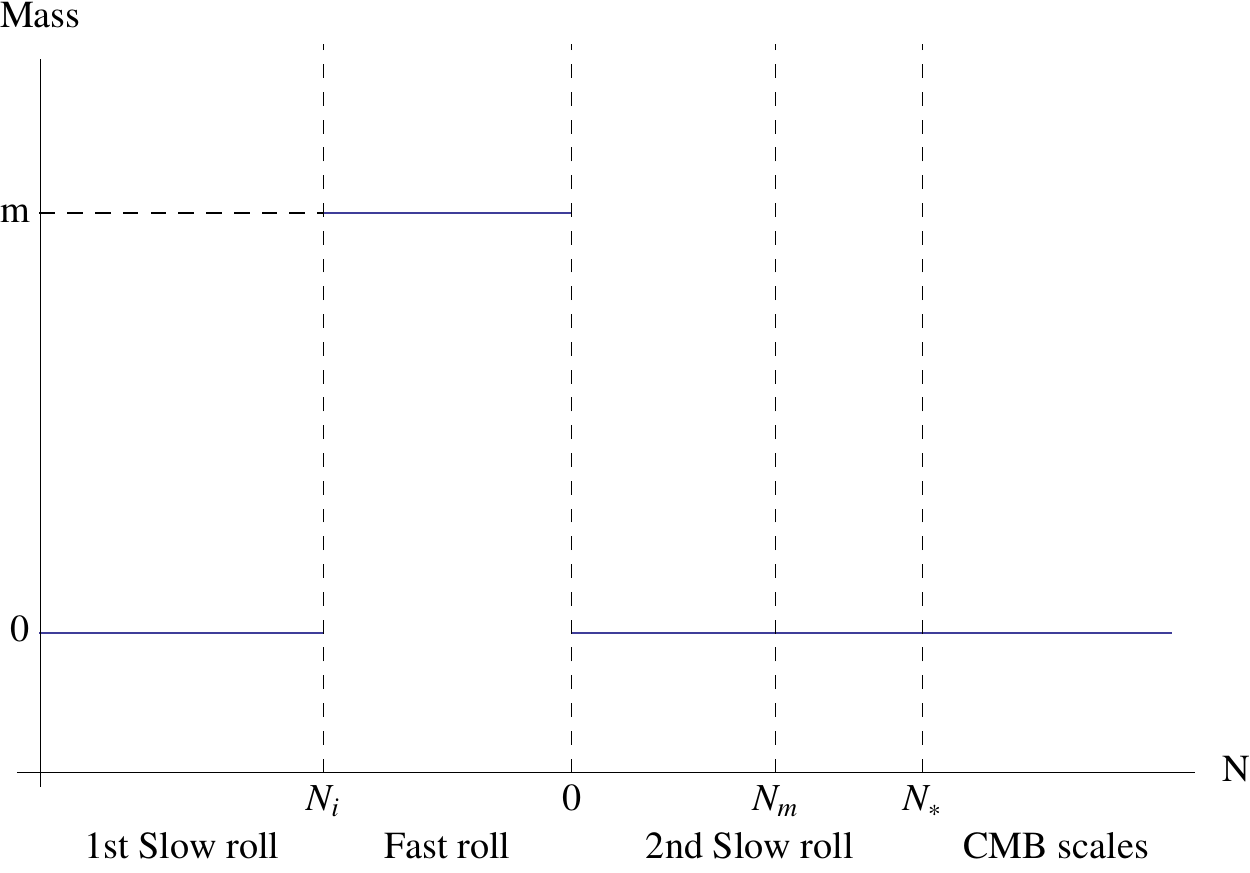}{The hierarchy of scales, and the timeline of the tachyonic fast roll scenario. The tachyonic fast roll phase lasts during $N_i<N<0$.}{0.7}{}
Then we will get a typical timeline as in \refig{Timeline}. We choose $a=1$ i.e.\ $N=0$ for the end of the fast roll phase, and assume its beginning lies at $N_i<0$. The mass scale is indicated as $N_m>0$ in \refig{Timeline}. To prevent the fast roll phase from affecting the CMB spectrum, we need the pivot scale to satisfy $N_*\ge N_m$, as we will show shortly.

\subsection{Enhancing the perturbations\label{ssec-Comparative enhancing rate}}
We focus on the fast roll phase and follow the steps of \refsec{Slow roll scenario}. The equation of motion for the perturbation $\delta\sigma$ in momentum space during the fast roll phase then becomes
\eq{\ddot{\delta\sigma}\vectf k+3H\dot{\delta\sigma}\vectf k+\left(\frac{k^2}{a^2}-m^2\right)\delta\sigma\vectf k=0.\label{eq-tc-dsk}}
Similar calculations will yield\footnote{
Here we only inspect the simplest setup, i.e.\ the tachyonic mass $m$ is switched on and off instantly, and remains constant during the period. In more realistic scenarios, the mass $m$ may depend on time, by writing $m(N)$ or $N_m(N)$. The subsequent calculations can still be performed as long as the WKB approximation is applicable. Although the specific results will depend on $N_m(N)$, the general consequence will not change, i.e.\ the field perturbations $\delta\sigma\vectf k$ will be enhanced in a scale-dependent way.}
\eq{\alpha\vectf k(N)=\left\{\begin{array}{l@{\hspace{0.5in}\mathrm{for\ }}l}
1,&N_k\ge N+N_m,\\
\frac{3}{2}-\sqrt{\frac{1}{4}+2\left(e^{2N_m}-e^{2(N_k-N)}\right)},&N_k<N+N_m.
\end{array}\right.\label{eq-tc-as}}

By comparing the values of $\alpha\vectf k(N)$ for the two scenarios, \refeq{ml-as2} and \refeq{tc-as}, we see that the outcome is different. Although the sub-Hubble modes are always redshifted with the rate $\alpha\vectf k=1$ in both the cases, the sub/super-Hubble boundary is shifted to $N_k=N_m+N$ in the fast roll scenario. Therefore the modes within $N<N_k<N_m+N$, which would be regarded as sub-Hubble and redshifted as $\alpha\vectf k=1$ in the slow roll scenario, are now in the fast roll scenario super-Hubble, with a lower damping rate $\alpha\vectf k<1$. In this sense, when the tachyonic mass is present, the small scale modes ($N_k>N+N_m$) are unaffected (with the same damping rate $\alpha\vectf k=1$), while the intermediate scales ($N<N_k<N+N_m$) are enhanced compared to the slow roll case.

The large scale modes ($N_k<N$) are also enhanced by the tachyonic mass comparatively, which is easy to see. Moreover, in the slow roll scenario we would always get $0<\alpha\vectf k<1$, but in the tachyonic fast roll scenario, it is possible to achieve $\alpha\vectf k<0$ when $N_m$ is sufficiently large, i.e.\ $m\gg H$. This corresponds to the case when the amplification by the tachyonic potential overcomes the expansion of the universe and the spatial inhomogeneities.

\fig{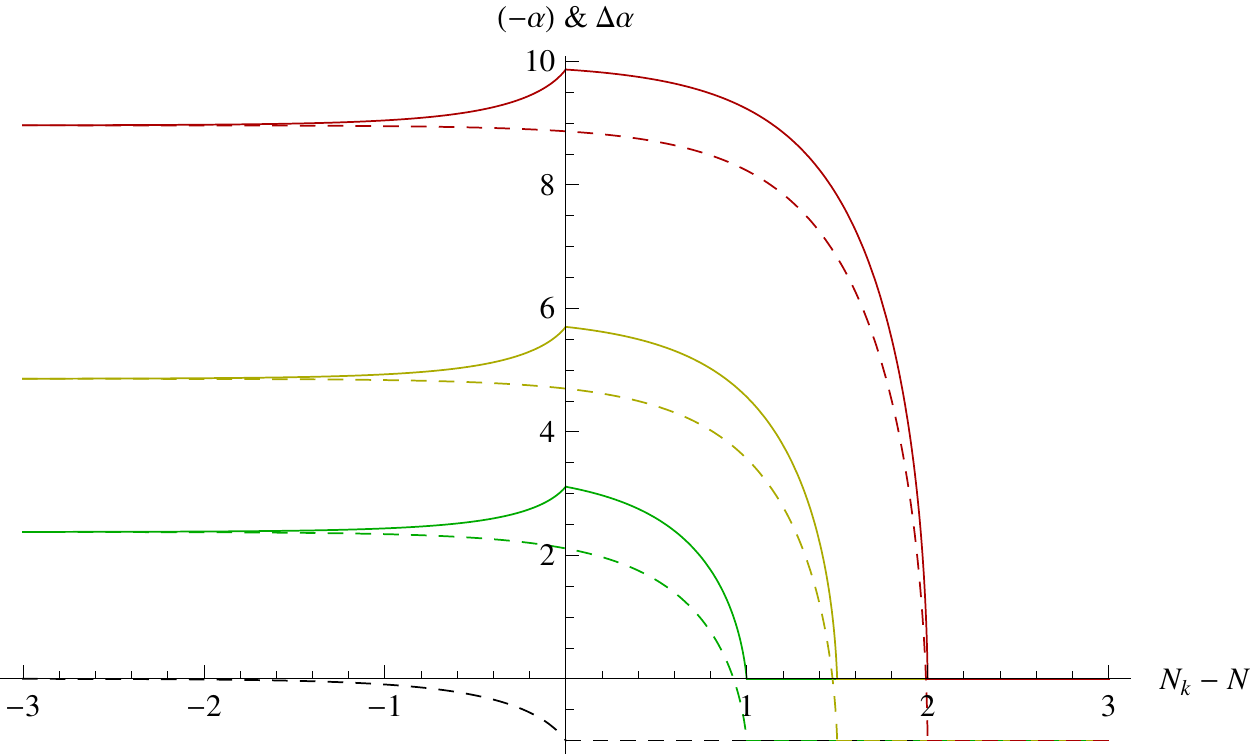}{A demonstration of $-\alpha\vectf k(N)$ and the \emph{enhance rate} $\Delta\alpha\vectf k(N)$ for some typical values of $N_m$. From bottom to top, black, green, yellow and red correspond to $N_m=0,1,1.5,2$ respectively. Dashed curves represent $-\alpha\vectf k(N)$, and solid ones represent $\Delta\alpha\vectf k(N)$.}{0.8}{}
We can then write the \emph{enhance rate} for the field perturbation $\delta\sigma\vectf k$, as the difference between the two damping rates,
\eq{\Delta\alpha\vectf k(N)\equiv\alpha\vectf k(N)\ml-\alpha\vectf k(N).}
Therefore we get the enhance rate for small, intermediate and large scales as
\eq{\Delta\alpha\vectf k(N)=\left\{\begin{array}{l@{\hspace{0.35in}\mathrm{for\ }}l}
0,&N_k\ge N+N_m,\\
\sqrt{\frac{1}{4}+2\left(e^{2N_m}-e^{2(N_k-N)}\right)}-\frac{1}{2},&N+N_m>N_k\ge N,\\
\sqrt{\frac{1}{4}+2\left(e^{2N_m}-e^{2(N_k-N)}\right)}-\sqrt{\frac{9}{4}-2e^{2(N_k-N)}},&N_k<N.
\end{array}\right.\label{eq-tc-Da}}
This is plotted with some typical values of $N_m$ in \refig{alpha}, from which we can see the enhancement can be quite significant $\Delta\alpha\vectf k\sim e^{N_m}$. Also, from \refeq{tc-Da} we can see that the small scales with $N_k\ge N+N_m$ are not affected by the fast roll phase. Remembering the fast roll phase lasts from $N=N_i<0$ to $N=0$, the scales $N_k\ge N_m$ will be totally unaffected, which is where we want the pivot scale to lie ($N_*\ge N_m$).

\ssecs{Generating CMB dipole asymmetry}
The relative enhancement in the fast roll scenario contributes an additional factor given by
\eq{\frac{\Delta\sigma^2}{P_{\delta\sigma_*}}=\int_{-\infty}^{N_m}4e^{2(N_k-N_*)+2\int_{N_i}^0\Delta\alpha\vectf k(N)\dd N}\dd N_k.\label{eq-tc-Ds2}}
Here we have neglected the integral region $N_m<N_k<N_*$, because this region is not enhanced by the tachyonic fast roll scenario, and has been shown in \refsec{Slow roll scenario} to generate only a negligible CMB asymmetry. The inner integral of $\Delta\alpha\vectf k$ is performed for the fast roll phase $N_i<N<0$. Then \refeq{tc-Ds2} can be recast into
\eq{\frac{\Delta\sigma^2}{P_{\delta\sigma_*}}=4e^{2(N_m-N_*)}\int_{-\infty}^{N_m}e^{2\beta(N_k)}\dd N_k,\label{eq-tc-Ds22}}
where we have defined
\eq{\beta(N_k)\equiv N_k-N_m+\int_{N_i}^0\Delta\alpha_k(N)\dd N.\label{eq-tc-b}}
Since the mode dependence in \refeq{tc-Ds22} have been absorbed into $\beta(N_k)$, the scale $k\subs{max}$ with the largest $\beta(N_k)$ will contribute most to the CMB asymmetry. The overall exponential coefficient in \refeq{tc-Ds22} simply means that a longer second slow roll phase, which will stretch the initial perturbation modes, leads to a weaker CMB asymmetry.

\fig{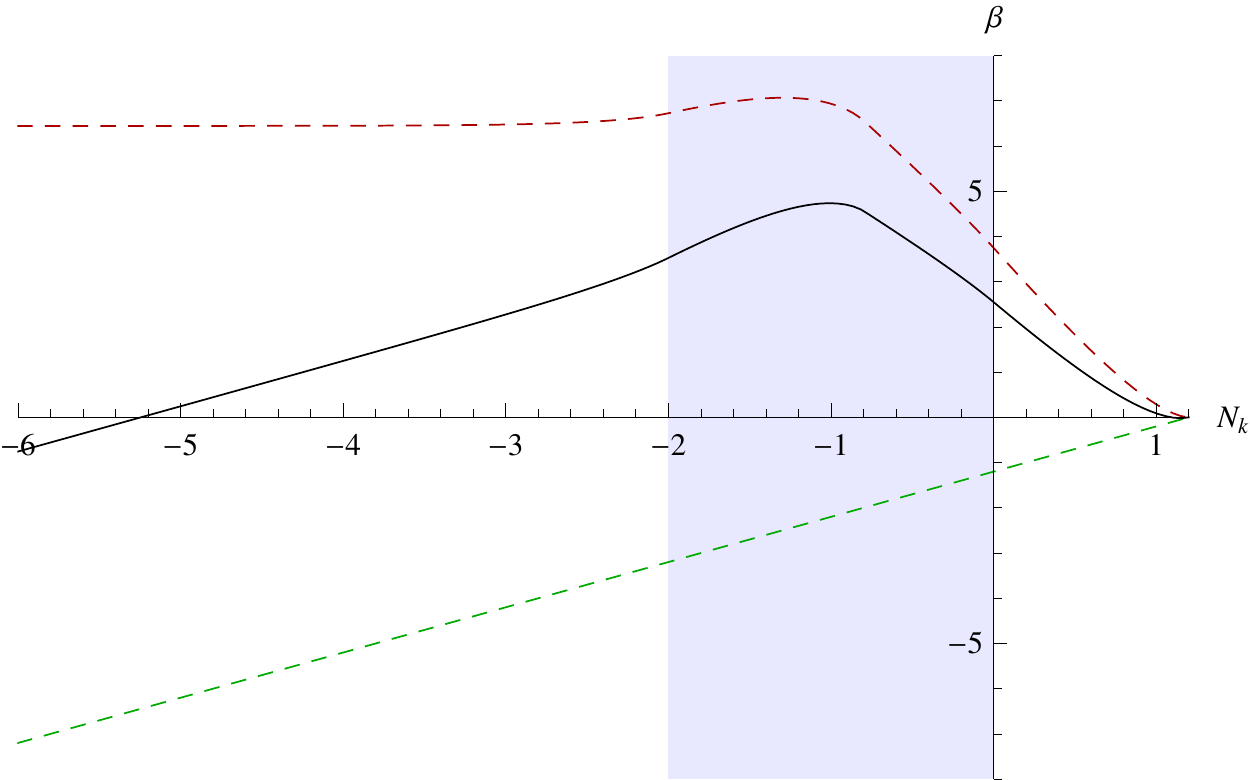}{The enhancement $\beta$ is shown in the black solid curve, where its components $N_k-N_m$ and the integral in \refeq{tc-b} are shown in green and red dashed curves respectively. The blue shaded region is the number of e-folds during the fast roll phase. We have taken the parameter values $N_i=-2, N_m=1.2, N_*=2.2$.}{0.8}{}

We would like $\beta(N_k)$ to peak at some scale $N_{k\subs{max}}$, or otherwise it is difficult to produce sufficient CMB dipole asymmetry. The peak mode $N_{k\subs{max}}$ can be solved from $\partial\beta(N_k)/\partial N_k=0$. Noticing $\Delta\alpha\vectf k(N)$ is only a function of $N-N_k$, this yields
\eq{\Delta\alpha_{k\subs{max}}(0)=\Delta\alpha_{k\subs{max}}(N_i)+1.\label{eq-sm-beq}}
Therefore there would be no peak if $\Delta\alpha_{k}(N)$ is always less than 1. The above condition requires the tachyonic mass to be heavy enough. According to \refeq{tc-Da}, we obtain
\eq{m^2\ge2H^2.}

The contribution to the CMB asymmetry would then mostly come from around the peak scale $N_{k\subs{max}}$. If we know the full width at half maximum (FWHM) of the peak, namely $\Delta N$, we can have a good estimation for the integral, hence writing \refeq{tc-Ds22} as
\eq{\frac{\Delta\sigma^2}{P_{\delta\sigma_*}}\approx4\Delta Ne^{2(N_m-N_*+\beta(N_{k\subs{max}}))}.\label{eq-tc-Ds23}}

A typical example of $\beta(N_k)$ is shown in \refig{beta}, for $N_i=-2$ and $N_m=1.2$, in which $\beta(N_k)$ peaks at about $N_k\approx-1$, with $\beta(N_{k\subs{max}})\approx4.7$. The half maximum lies at $\beta(N_{k\subs{max}})-\frac{1}{2}\ln2$ with $\Delta N\approx0.8$. We do not want the pivot scale spectrum to be modified by the fast roll phase, so we yield $N_*>N_m$, and therefore, we take $N_*=N_m+1$. Plugging these numbers into \refeq{tc-Ds23} will give
\eq{\frac{|\Delta\sigma|}{\sqrt{P_{\delta\sigma_*}}}=72.\label{eq-tc-Dsn}}
This result satisfies the necessary condition \refeq{ca-Dslim}, but not \refeq{ca-Dslims}. From this we can see that the bispectrum cancellation can indeed enhance the CMB dipole asymmetry, but this would require another source for the curvature perturbations. However, with a different choice of the parameters $N_i=-1.8$, $N_m=1.4$ and $N_*=2.4$, it is still possible to satisfy \refeq{ca-Dslims}, as shown in \refsec{A viable model}. Also, \refeq{tc-Ds23} and \refeq{tc-Dsn} gives the average CMB asymmetry on any direction. When the direction with the strongest asymmetry is chosen, the CMB asymmetry would be larger.

We also have to make sure that the perturbations remain small throughout the dynamics. This typically requires the curvature perturbations generated by the $\sigma$ field to have a power spectrum $P_{\zeta_{\delta\sigma_k}}<1$. Since there can be other sources of curvature perturbations, we define a ratio for $\sigma$ at the pivot scale
\eq{R^2\equiv\frac{P_{\zeta_{\delta\sigma_*}}}{P_{\zeta}}\le1.\label{ratio-R}}
The constraint $P_{\zeta_{\delta\sigma_*}}<1$ then becomes
\eqa{P_{\zeta_{\delta\sigma_k}}&=&\left.P_{\zeta_{\delta\sigma_k}}\right\ml e^{2\int_{N_i}^0\Delta\alpha_k(N)\dd N}\nonumber\\
&=&P_{\zeta}R^2e^{2\int_{N_i}^0\Delta\alpha_k(N)\dd N}<1,\label{eq-kc-pze}}
where we have used $P_{\zeta_{\delta\sigma_k}}\ml=P_{\zeta_{\delta\sigma_*}}$ for any mode $k$. This constrains the total amount of enhancement, i.e.\ the height of the red dashed curve in \refig{beta},  by
\eq{\int_{N_i}^0\Delta\alpha_k(N)\dd N<-\frac{1}{2}\ln P_\zeta-\ln R,\hspace{0.3in}\mathrm{for\ any\ }N_k<0.\label{eq-kc-pac}}

Because $R\le1$ and $P_\zeta=2.196\times10^{-9}$, in the example, \refig{beta}, it is easy to see that the red curve is lower than $-\frac{1}{2}\ln P_\zeta\approx10$, and therefore the condition \refeq{kc-pac} is well satisfied.

\secs{Known constraints}
\ssecs{Quasars}
\fig{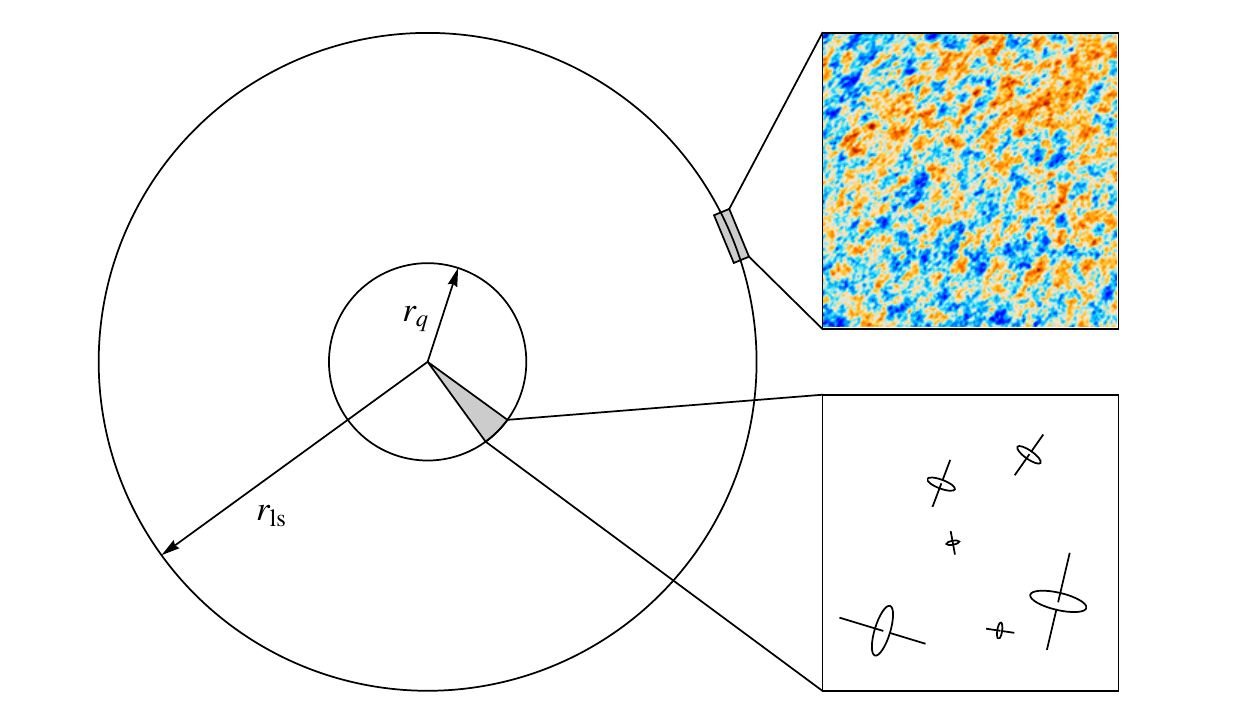}{A schematic figure on the dipole asymmetries of the CMB and quasars. The outer sphere is the LSS and the inner one contains all the observed quasars. Therefore the quasar observations can only constrain the asymmetry in the distance scales smaller than $r_q$, while the asymmetry in the distance scale $r_{ls}$ can be much larger. In this sense, we need the running in the asymmetry factor $A$ in the length scale in order to satisfy the quasar constraint, which has been observed in the Planck observations in figure 28 of \cite{Ade:2013nlj}, though in $l$ space.}{0.8}{}
The quasar observations \cite{Hirata:2009ar} constrain the universe asymmetry in the quasar scale $N_q>N_*$. If we define $r_q$ as our distance to the farthest quasar, we can write $N_q$, the length scale of our distance to the quasars, similar with \refeq{ml-rls}, as (see also \refig{CMBvq})
\eq{r_q=\frac{e^{-N_q}}{H}.}

The quasar observations find no asymmetry, requiring $A<0.02$ in the quasar scale $N_q$; see \cite{Hirata:2009ar}. From \refeq{ca-Aeqs} and \refeq{ca-Aup}, we find that it can be accommodated with the CMB scale asymmetry $A\sim0.07$, if the non-Gaussianity parameter, $\fNL$ or $\tNL$, has a running. The running should be strong enough to reduce the non-Gaussianity to $\lesssim1/4$ during inflation, from $N=N_*$ to $N=N_q$.

The amount of non-Gaussianity and its running depend very much on the model, but many existing models can provide such a running. In the spectator model we consider in \cite{Wang:2013oea}, this can be achieved if the effective mass of the spectator field runs between the Hubble exits of the pivot and quasar scales.

\ssecs{Quadrupole and octupole}
The source of CMB asymmetry should not generate excessive quadrupole or octupole in the CMB. Following the conventions in \cite{Erickcek:2008jp,Erickcek:2008sm}, we replicate their derived constraints here, from eq.\ (4), and eq.\ (5) of Ref.~\cite{Erickcek:2008sm}
\eqa{(kx_d)^2|\Phi_{\vec k}(\tau_d)\sin\overline{\omega}|&\lesssim&5.8{\cal Q},\hspace{0.3in}\mathrm{for\ quadrupole,}\\
(kx_d)^3|\Phi_{\vec k}(\tau_d)\cos\overline{\omega}|&\lesssim&32{\cal O},\hspace{0.34in}\mathrm{for\ octupole,}}
where ${\cal Q}=1.8\times10^{-5}$ and ${\cal O}=2.7\times10^{-5}$. We rephrase them with our convention, with $k x_d|\Phi_{k}(\tau_d)|=\frac{1}{3}|\Delta\zeta|=\sqrt{P_\zeta}\frac{|\Delta\sigma|}{3\sqrt{P_{\delta\sigma_*}}}$ where $k x_d=\sqrt2e^{N_{k\subs{max}}-N_*}$. After neglecting the $\sin$ and $\cos$ functions, these two inequalities become
\eqa{N_{k\subs{max}}-N_*&\lesssim&\ln\frac{17.4{\cal Q}}{\sqrt{2P_\zeta}}\frac{\sqrt{P_{\delta\sigma_*}}}{|\Delta\sigma|},\hspace{0.355in}\mathrm{for\ quadrupole},\\
N_{k\subs{max}}-N_*&\lesssim&\frac{1}{2}\ln\frac{48{\cal O}}{\sqrt{P_\zeta}}\frac{\sqrt{P_{\delta\sigma_*}}}{|\Delta\sigma|},\hspace{0.3in}\mathrm{for\ octupole}.}

Therefore the quadrupole and octupole constraints put a lower bound on $N_*$, the e-folding of the second slow roll phase, see \refig{Timeline}. In the example shown in \refig{beta}, we have $N_{k\subs{max}}\approx-1$. By plugging in the values of ${\cal O,\ Q},\ P_\zeta,\ N_*$ and $|\Delta\sigma|/\sqrt{P_{\delta\sigma_*}}$ from the example \refig{beta} and \refeq{tc-Dsn}, we can see that both the quadrupole and octupole constraints are satisfied.

\secs{A viable model}
As we have seen in previous sections that the quasar constraint can always be satisfied with a proper running non-Gaussianity, \refeq{kc-pac} constrains the amount of total enhancement, and the quadrupole and octupole constrain the length of the second slow roll phase. To achieve the maximum CMB dipole asymmetry $A$, we need to maximize the total enhancement and minimize the length of the second slow roll phase. We examine the maximum possible value for $A$ in the following two models mentioned in \refsec{CMB dipole asymmetry}.

\fig{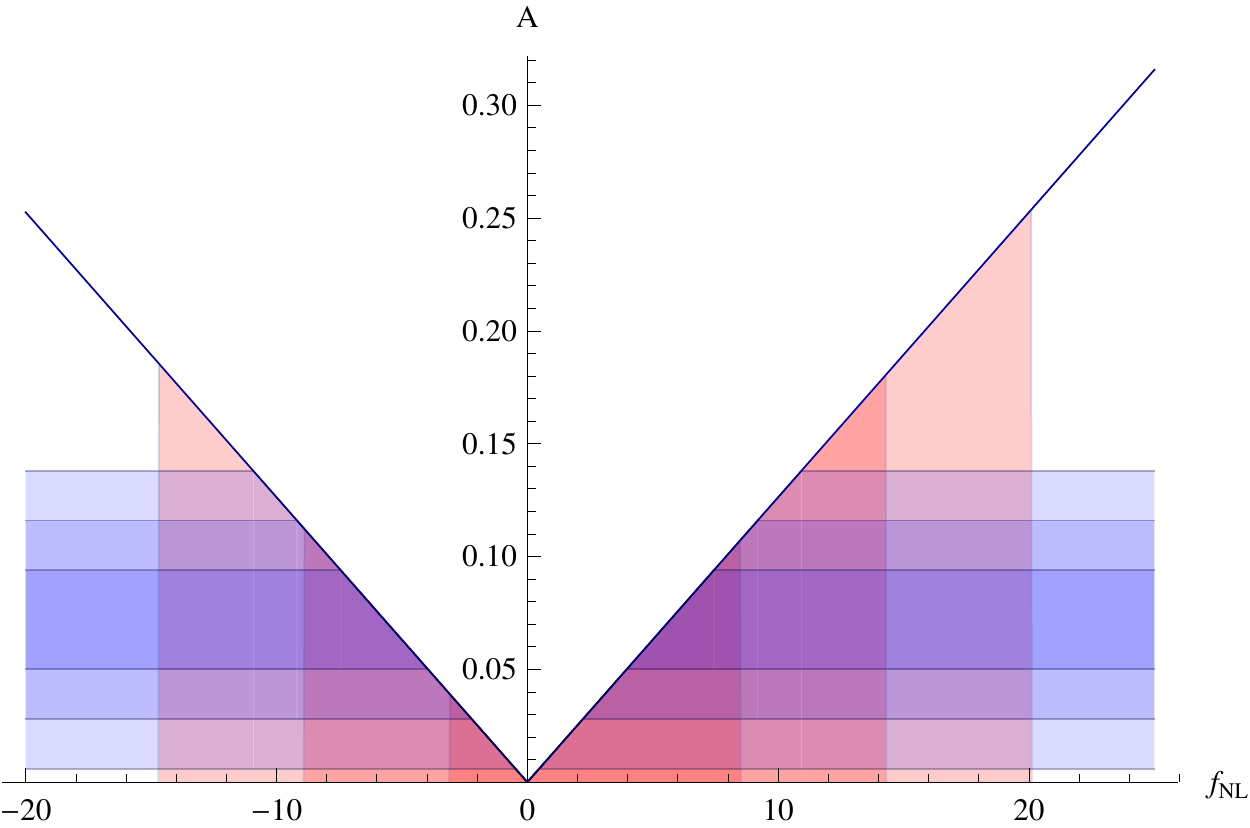}{The parameter plot for the inflaton and spectator model in \refsec{A viable model}. The blue line shows the maximum CMB dipole asymmetry can be reached by any bispectrum $\fNL$. The vertical red bands indicate the latest Planck observational bounds for the local bispectrum $\fNL$, for the $1\sigma$, $2\sigma$ and $3\sigma$ regions. The horizontal blue bands indicate the Planck observational bounds for the CMB dipole asymmetry, also for the $1\sigma$, $2\sigma$ and $3\sigma$ regions. The neighboring area of $\fNL\approx 7$ and $A\approx0.07$ is within $1\sigma$ C.L. for both the observables.}{0.8}{}

\begin{itemize}
\item {\bf Inflaton, and the spectator}: Inflation is driven by the $\phi$ field, while the curvature perturbations come from $\sigma$, which is a spectator field. The $\sigma$ field generates mild local bispectrum $\fNL=2.7\pm11.6$ (at $2\sigma$) \cite{Ade:2013ydc}. Combining the above constraints gives us an upper bound for the CMB dipole asymmetry
\eq{A<1.2\sqrt{8.7{\cal Q}/\sqrt2}\Delta N^\frac{1}{4}|\fNL|,\label{eq-vm-twof}}
where we have used the quadrupole constraint, but neglected the octupole because it is much weaker.

After putting in the values for $Q\lesssim1.8\times10^{-5}$, $\Delta N\approx1$ and $\fNL<14.3$, we can find the maximum value of $A\lesssim0.18$, allowed by the constraints. A specific example for this model can be found when $N_i=-1.8$, $N_m=1.4$, $N_*=2.4$, and $\fNL=7$, which will generate $A=0.07$. Also, this model can be excluded for $A=0.07$ if the future experiments improve the errors on $\fNL$, so that $|\fNL|\lesssim4$. The parameter plot for this model based on \refeq{vm-twof} is shown in \refig{twof}.

\item {\bf Inflaton, spectator, and the other}: Another possibility is an additional field $\chi$ also contributes to the power spectrum of the curvature perturbations, so $\sigma$  will only contribute the amount $P_{\zeta_{\delta\sigma_*}}=R^2P_\zeta$ where $R<1$. In this case, each of them can generate positive or negative $\fNL$, but in all they cancel. It will produce a stronger dipole asymmetry than the previous case, giving
\eq{A<\sqrt{8.7{\cal Q}/\sqrt2}\Delta N^\frac{1}{4}R^{-\frac{1}{2}}\sqrt{\tNL}.}
We substitute the values of $Q\lesssim1.8\times10^{-5}$, $\Delta N\approx1$ and $\tNL<2800$, so the above equation yields $A\lesssim0.56R^{-\frac{1}{2}}$. Therefore $A=0.07$ is also allowed, and this model can generate a much stronger dipole asymmetry than the previous one.
\end{itemize}

\secs{Conclusion}
In this paper, we have shown that a spectator field can indeed generate the observed CMB dipole asymmetry. However it requires a brief phase of fast roll before the Hubble exit of the relevant scales to amplify its perturbations. We have shown that this primordial mechanism explains the CMB dipole asymmetry while satisfying {\it all } known constraints, the quasar, quadrupole and octupole constraints.

In the inflaton-spectator two-field setup, this mechanism can generate a maximum of $A\lesssim0.18$, under the current constraint on $\fNL=2.7\pm 11.6$ (at $2\sigma$). This can be further constrained by future observations, for instance more precise value on $\fNL$. In the three-field setup, the CMB dipole asymmetry $A$ can be  enhanced by $\tNL$ due to the fine cancellation between the $\pm\fNL$ sources. As a result the asymmetry can be much larger, i.e.  $A\lesssim0.56R^{-1}$ for $\tNL<2800$, where $R$ is defined in Eq.~(\ref{ratio-R}).

In the specific examples, we have shown that the CMB dipole asymmetry can still be explained well within an inflation paradigm, but it will require more than one field. We would need a spectator field to generate the curvature perturbations. Single-field inflation is however difficult to generate the observed CMB dipole asymmetry, because of its lack of non-Gaussianity in its simplest form.
 
\acknowledgments{}
We would like to thank Christopher Hirata, Matthew Kleban, David Lyth, John McDonald and Yun-Song Piao for helpful discussions. AM is supported by the Lancaster-Manchester-Sheffield Consortium for Fundamental Physics under STFC grant ST/J000418/1.
\bibliographystyle{jcap}
\bibliography{Main}

\providecommand{\href}[2]{#2}\begingroup\raggedright\begin{thebibliography}{10}

\bibitem{Bennett:1996ce}
C.~Bennett, A.~Banday, K.~Gorski, G.~Hinshaw, P.~Jackson, {\em et.~al.}, {\it
  {Four year COBE DMR cosmic microwave background observations: Maps and basic
  results}},  {\em Astrophys.J.} {\bf 464} (1996) L1--L4,
  [\href{http://xxx.lanl.gov/abs/astro-ph/9601067}{{\tt astro-ph/9601067}}].

\bibitem{Hinshaw:2012fq}
G.~Hinshaw, D.~Larson, E.~Komatsu, D.~Spergel, C.~Bennett, {\em et.~al.}, {\it
  {Nine-Year Wilkinson Microwave Anisotropy Probe (WMAP) Observations:
  Cosmological Parameter Results}},
  \href{http://xxx.lanl.gov/abs/1212.5226}{{\tt arXiv:1212.5226}}.

\bibitem{Ade:2013zuv}
{\bf Planck Collaboration} Collaboration, P.~Ade {\em et.~al.}, {\it {Planck
  2013 results. XVI. Cosmological parameters}},
  \href{http://xxx.lanl.gov/abs/1303.5076}{{\tt arXiv:1303.5076}}.

\bibitem{Eriksen:2003db}
H.~Eriksen, F.~Hansen, A.~Banday, K.~Gorski, and P.~Lilje, {\it {Asymmetries in
  the Cosmic Microwave Background anisotropy field}},  {\em Astrophys.J.} {\bf
  605} (2004) 14--20, [\href{http://xxx.lanl.gov/abs/astro-ph/0307507}{{\tt
  astro-ph/0307507}}].

\bibitem{Eriksen:2007pc}
H.~K. Eriksen, A.~Banday, K.~Gorski, F.~Hansen, and P.~Lilje, {\it
  {Hemispherical power asymmetry in the three-year Wilkinson Microwave
  Anisotropy Probe sky maps}},  {\em Astrophys.J.} {\bf 660} (2007) L81--L84,
  [\href{http://xxx.lanl.gov/abs/astro-ph/0701089}{{\tt astro-ph/0701089}}].

\bibitem{Ade:2013nlj}
{\bf Planck Collaboration} Collaboration, P.~Ade {\em et.~al.}, {\it {Planck
  2013 results. XXIII. Isotropy and Statistics of the CMB}},
  \href{http://xxx.lanl.gov/abs/1303.5083}{{\tt arXiv:1303.5083}}.

\bibitem{Erickcek:2008sm}
A.~L. Erickcek, M.~Kamionkowski, and S.~M. Carroll, {\it {A Hemispherical Power
  Asymmetry from Inflation}},  {\em Phys.Rev.} {\bf D78} (2008) 123520,
  [\href{http://xxx.lanl.gov/abs/0806.0377}{{\tt arXiv:0806.0377}}].

\bibitem{Erickcek:2009at}
A.~L. Erickcek, C.~M. Hirata, and M.~Kamionkowski, {\it {A Scale-Dependent
  Power Asymmetry from Isocurvature Perturbations}},  {\em Phys.Rev.} {\bf D80}
  (2009) 083507, [\href{http://xxx.lanl.gov/abs/0907.0705}{{\tt
  arXiv:0907.0705}}].

\bibitem{Ade:2013uln}
{\bf Planck Collaboration} Collaboration, P.~Ade {\em et.~al.}, {\it {Planck
  2013 results. XXII. Constraints on inflation}},
  \href{http://xxx.lanl.gov/abs/1303.5082}{{\tt arXiv:1303.5082}}.

\bibitem{Wang:2013hva}
L.~Wang, E.~Pukartas, and A.~Mazumdar, {\it {Visible sector inflation and the
  right thermal history in light of Planck data}},
  \href{http://xxx.lanl.gov/abs/1303.5351}{{\tt arXiv:1303.5351}}.

\bibitem{Mazumdar:2010sa}
A.~Mazumdar and J.~Rocher, {\it {Particle physics models of inflation and
  curvaton scenarios}},  {\em Phys.Rept.} {\bf 497} (2011) 85--215,
  [\href{http://xxx.lanl.gov/abs/1001.0993}{{\tt arXiv:1001.0993}}].

\bibitem{Allahverdi:2006iq}
R.~Allahverdi, K.~Enqvist, J.~Garcia-Bellido, and A.~Mazumdar, {\it {Gauge
  invariant MSSM inflaton}},  {\em Phys.Rev.Lett.} {\bf 97} (2006) 191304,
  [\href{http://xxx.lanl.gov/abs/hep-ph/0605035}{{\tt hep-ph/0605035}}].

\bibitem{Allahverdi:2006we}
R.~Allahverdi, K.~Enqvist, J.~Garcia-Bellido, A.~Jokinen, and A.~Mazumdar, {\it
  {MSSM flat direction inflation: Slow roll, stability, fine tunning and
  reheating}},  {\em JCAP} {\bf 0706} (2007) 019,
  [\href{http://xxx.lanl.gov/abs/hep-ph/0610134}{{\tt hep-ph/0610134}}].

\bibitem{Allahverdi:2006cx}
R.~Allahverdi, A.~Kusenko, and A.~Mazumdar, {\it {A-term inflation and the
  smallness of neutrino masses}},  {\em JCAP} {\bf 0707} (2007) 018,
  [\href{http://xxx.lanl.gov/abs/hep-ph/0608138}{{\tt hep-ph/0608138}}].

\bibitem{Choudhury:2013jya}
S.~Choudhury, A.~Mazumdar, and S.~Pal, {\it {Low and High scale MSSM inflation,
  gravitational waves and constraints from Planck}},
  \href{http://xxx.lanl.gov/abs/1305.6398}{{\tt arXiv:1305.6398}}.

\bibitem{Wang:2013lda}
L.~Wang and A.~Mazumdar, {\it {Small non-Gaussianity and dipole asymmetry in
  the CMB}},  \href{http://xxx.lanl.gov/abs/1304.6399}{{\tt arXiv:1304.6399}}.

\bibitem{Dai:2013kfa}
L.~Dai, D.~Jeong, M.~Kamionkowski, and J.~Chluba, {\it {The Pesky Power
  Asymmetry}},  \href{http://xxx.lanl.gov/abs/1303.6949}{{\tt
  arXiv:1303.6949}}.

\bibitem{Lyth:2013vha}
D.~H. Lyth, {\it {The CMB asymmetry from inflation}},
  \href{http://xxx.lanl.gov/abs/1304.1270}{{\tt arXiv:1304.1270}}.

\bibitem{Chang:2013vla}
Z.~Chang and S.~Wang, {\it {Inflation and primordial power spectra at
  anisotropic spacetime inspired by Planck's constraints on isotropy of CMB}},
  \href{http://xxx.lanl.gov/abs/1303.6058}{{\tt arXiv:1303.6058}}.

\bibitem{McDonald:2013aca}
J.~McDonald, {\it {Isocurvature and Curvaton Perturbations with Red Power
  Spectrum and Large Hemispherical Asymmetry}},
  \href{http://xxx.lanl.gov/abs/1305.0525}{{\tt arXiv:1305.0525}}.

\bibitem{Namjoo:2013fka}
M.~H. Namjoo, S.~Baghram, and H.~Firouzjahi, {\it {Hemispherical Asymmetry and
  Local non-Gaussianity: a Consistency Condition}},
  \href{http://xxx.lanl.gov/abs/1305.0813}{{\tt arXiv:1305.0813}}.

\bibitem{Zhao:2013jya}
W.~Zhao, {\it {Directional dependence of CMB parity asymmetry}},
  \href{http://xxx.lanl.gov/abs/1306.0955}{{\tt arXiv:1306.0955}}.

\bibitem{Donoghue:2007ze}
J.~F. Donoghue, K.~Dutta, and A.~Ross, {\it {Non-isotropy in the CMB power
  spectrum in single field inflation}},  {\em Phys.Rev.} {\bf D80} (2009)
  023526, [\href{http://xxx.lanl.gov/abs/astro-ph/0703455}{{\tt
  astro-ph/0703455}}].

\bibitem{Chen:2013eaa}
X.~Chen and Y.~Wang, {\it {Relic Vector Field and CMB Large Scale Anomalies}},
  \href{http://xxx.lanl.gov/abs/1305.4794}{{\tt arXiv:1305.4794}}.

\bibitem{Liu:2013kea}
Z.-G. Liu, Z.-K. Guo, and Y.-S. Piao, {\it {Obtaining the CMB anomalies with a
  bounce from the contracting phase to inflation}},
  \href{http://xxx.lanl.gov/abs/1304.6527}{{\tt arXiv:1304.6527}}.

\bibitem{Linde:2001ae}
A.~D. Linde, {\it {Fast roll inflation}},  {\em JHEP} {\bf 0111} (2001) 052,
  [\href{http://xxx.lanl.gov/abs/hep-th/0110195}{{\tt hep-th/0110195}}].

\bibitem{Mazumdar:2012rs}
A.~Mazumdar and L.~Wang, {\it {Creating perturbations from a decaying field
  during inflation}},  {\em Phys.Rev.} {\bf D87} (2013) 083501,
  [\href{http://xxx.lanl.gov/abs/1210.7818}{{\tt arXiv:1210.7818}}].

\bibitem{Wang:2013oea}
L.~Wang and A.~Mazumdar, {\it {Cosmological perturbations from a Spectator
  field during inflation}},  {\em JCAP} {\bf 1305} (2013) 012,
  [\href{http://xxx.lanl.gov/abs/1302.2637}{{\tt arXiv:1302.2637}}].

\bibitem{Moroi:2001ct}
T.~Moroi and T.~Takahashi, {\it {Effects of cosmological moduli fields on
  cosmic microwave background}},  {\em Phys.Lett.} {\bf B522} (2001) 215--221,
  [\href{http://xxx.lanl.gov/abs/hep-ph/0110096}{{\tt hep-ph/0110096}}].

\bibitem{Lyth:2001nq}
D.~H. Lyth and D.~Wands, {\it {Generating the curvature perturbation without an
  inflaton}},  {\em Phys.Lett.} {\bf B524} (2002) 5--14,
  [\href{http://xxx.lanl.gov/abs/hep-ph/0110002}{{\tt hep-ph/0110002}}].

\bibitem{Enqvist:2001zp}
K.~Enqvist and M.~S. Sloth, {\it {Adiabatic CMB perturbations in pre - big bang
  string cosmology}},  {\em Nucl.Phys.} {\bf B626} (2002) 395--409,
  [\href{http://xxx.lanl.gov/abs/hep-ph/0109214}{{\tt hep-ph/0109214}}].

\bibitem{Lyth:2002my}
D.~H. Lyth, C.~Ungarelli, and D.~Wands, {\it {The Primordial density
  perturbation in the curvaton scenario}},  {\em Phys.Rev.} {\bf D67} (2003)
  023503, [\href{http://xxx.lanl.gov/abs/astro-ph/0208055}{{\tt
  astro-ph/0208055}}].

\bibitem{Mazumdar:2011xe}
A.~Mazumdar and S.~Nadathur, {\it {The curvaton scenario in the MSSM and
  predictions for non-Gaussianity}},  {\em Phys.Rev.Lett.} {\bf 108} (2012)
  111302, [\href{http://xxx.lanl.gov/abs/1107.4078}{{\tt arXiv:1107.4078}}].

\bibitem{Ade:2013ydc}
{\bf Planck Collaboration} Collaboration, P.~Ade {\em et.~al.}, {\it {Planck
  2013 Results. XXIV. Constraints on primordial non-Gaussianity}},
  \href{http://xxx.lanl.gov/abs/1303.5084}{{\tt arXiv:1303.5084}}.

\bibitem{Enqvist:2004ey}
K.~Enqvist, A.~Jokinen, A.~Mazumdar, T.~Multamaki, and A.~Vaihkonen, {\it
  {Non-Gaussianity from preheating}},  {\em Phys.Rev.Lett.} {\bf 94} (2005)
  161301, [\href{http://xxx.lanl.gov/abs/astro-ph/0411394}{{\tt
  astro-ph/0411394}}].

\bibitem{Enqvist:2005qu}
K.~Enqvist, A.~Jokinen, A.~Mazumdar, T.~Multamaki, and A.~Vaihkonen, {\it
  {Non-Gaussianity from instant and tachyonic preheating}},  {\em JCAP} {\bf
  0503} (2005) 010, [\href{http://xxx.lanl.gov/abs/hep-ph/0501076}{{\tt
  hep-ph/0501076}}].

\bibitem{Jokinen:2005by}
A.~Jokinen and A.~Mazumdar, {\it {Very large primordial non-gaussianity from
  multi-field: application to massless preheating}},  {\em JCAP} {\bf 0604}
  (2006) 003, [\href{http://xxx.lanl.gov/abs/astro-ph/0512368}{{\tt
  astro-ph/0512368}}].

\bibitem{Hirata:2009ar}
C.~M. Hirata, {\it {Constraints on cosmic hemispherical power anomalies from
  quasars}},  {\em JCAP} {\bf 0909} (2009) 011,
  [\href{http://xxx.lanl.gov/abs/0907.0703}{{\tt arXiv:0907.0703}}].

\bibitem{Erickcek:2008jp}
A.~L. Erickcek, S.~M. Carroll, and M.~Kamionkowski, {\it {Superhorizon
  Perturbations and the Cosmic Microwave Background}},  {\em Phys.Rev.} {\bf
  D78} (2008) 083012, [\href{http://xxx.lanl.gov/abs/0808.1570}{{\tt
  arXiv:0808.1570}}].

\end{thebibliography}\endgroup
\end{document}